\def\umass{1}
\def\itc{2}
\def\kavli{3}
\def\asu{4}
\def\alabama{5}
\def\germany{6}

\documentclass[twocolumn]{aastex6}
\usepackage{natbib}

\usepackage{graphicx, verbatim}
\usepackage[space]{grffile}
\usepackage{latexsym}
\usepackage{amsfonts,amsmath,amssymb}
\usepackage{url}
\usepackage{fancyref}

\begin{document}

\title{Constraining The Single-Degenerate Channel of Type Ia Supernovae With Stable Iron-Group Elements in SNR 3C 397}



\author{
Pranav Dave,\altaffilmark{\umass}
Rahul Kashyap,\altaffilmark{\umass}
Robert Fisher,\altaffilmark{\umass, \itc, \kavli}
Frank Timmes,\altaffilmark{\asu}
Dean Townsley,\altaffilmark{\alabama}
Chris Byrohl \altaffilmark{\germany}
}


\altaffiltext {\umass} {Department of Physics, University of Massachusetts Dartmouth, 285 Old Westport Road, North Dartmouth, MA 02740, USA}
\altaffiltext {\itc} {Institute for Theory and Computation, Harvard-Smithsonian Center for Astrophysics, 60 Garden Street, Cambridge, MA 02138, USA}
\altaffiltext {\kavli} {Kavli Institute for Theoretical Physics, Kohn Hall, University of California at Santa Barbara, Santa Barbara, CA 93106, USA}
\altaffiltext {\asu} {School of Earth and Space Exploration, Arizona State University, Tempe, AZ 85287, USA}
\altaffiltext {\alabama} {Department of Physics \& Astronomy, Box 870324, University of Alabama, Tuscaloosa, AL 35487, USA}
\altaffiltext{\germany}{Institut f{\"u}r Astrophysik, Georg August Universit{\"a}t G{\"o}ttingen, Friedrich-Hund-Platz 1, D-37077 G{\"o}ttingen, Germany}

\begin{abstract}
Recent Suzaku X-ray spectra of SNR 3C 397 indicate enhanced stable
iron-group element abundances of Ni, Mn, Cr, and Fe.  Seeking to
address key questions about the progenitor and explosion mechanism of 3C 397, we compute
nucleosynthetic yields from a suite of multidimensional hydrodynamics
models in the near-Chandrasekhar mass, single-degenerate paradigm for
supernova Type Ia.  Varying the progenitor white dwarf internal
structure, composition, ignition, and explosion mechanism, we find the
best match to the observed iron-peak elements of 3C 397 are dense
(central density $\ge$ 6$\times$10$^{9}$ g cm$^{-3}$), low-carbon white dwarfs that undergo a weak, centrally-ignited deflagration, followed by a subsequent
detonation. The amount of $^{56}$Ni produced is consistent with a normal
or bright normal supernova Type Ia. A pure deflagration of a centrally-ignited, low central density ($\simeq$ 2$\times$10$^{9}$ g cm$^{-3}$) progenitor white
dwarf, frequently considered in the literature, is also found to produce good agreement with 3C 397 nucleosynthetic yields, but leads to a subluminous SN Ia event, in conflict with X-ray linewidth data.  Additionally, in contrast to prior work which suggested a large super-solar metallicity for the white
dwarf progenitor for SNR 3C 397, we find satisfactory agreement for solar and sub-solar
metallicity progenitors. We discuss a range of implications our results
have for the single-degenerate channel.


\end{abstract}


\section{Introduction}
\label {introduction}

Type Ia supernovae (SNe Ia), are white dwarf stars (WDs) composed of carbon and oxygen which undergo explosive nuclear burning.  SNe are important across many astrophysical domains, serving as  sources of momentum and energy \citep {mckeeostriker77, lietal15},  turbulence \citep {elmegreenscalo04, joungmaclow06}, cosmic rays \citep {baadezwicky34, ginzburg64, blandfordeichler87, drury12}, and enriched isotopes for the interstellar medium \citep{kobayashietal06}. The class of SNe Ia are additionally crucially important as standardizable candles for cosmology \citep {phillips93},  and as  endpoints of low-mass binary stellar evolution. 

Since an isolated WD is inherently stable, virtually all SN Ia mechanisms\footnote {The one exception perhaps being  the pycnonuclear-driven model of \cite {chiosietal15}} have invoked the presence of one or two stellar companions. However, the nature of the stellar companion and the explosion mechanism have remained unclear. The two most frequently-discussed possibilities are for a companion degenerate WD in the double-degenerate (DD) channel \citep {webbink84, ibentutukov84} or another non-degenerate star in the single-degenerate (SD) channel \citep {whelaniben73}. \deleted {Additionally, a WD accreting helium may also produce a SN Ia through the double-detonation channel \citep {nomoto82}.} \added{Additionally, other possibilities include He-accreting WDs in the double-detonation channel \citep {nomoto82}, the merging of C/O WDs with asymptotic giant brianch stars in the core-degenerate scenario \citep {raskinetal10,ilkovsoker13}, and the collision of WDs in triple stellar systems \citep {kushniretal13}.}


In the SD channel, a WD accretes matter from a main sequence or red giant star, until the WD eventually ignites as a deflagration, which may subsequently lead to a detonation, and produce a Type Ia supernova. Under steady mass accretion, the WD builds up its mass until it approaches the Chandrasekhar limit, igniting as a near-Chandrasekhar (near-$M_\text{ch}$) WD \citep {townsleybildsten05}.

For decades, the SD channel was considered the leading model to explain the relative uniformity of SNe Ia properties -- see e.g. \citet {hoylefowler60}, \citet {nomotoetal84}, \citet {niemeyerwoosley97}, and \citet {hillenbrandtropke10}. Yet, with increasing observational evidence, the SD channel was found to be inconsistent with a range of observational constraints, including the delay-time distribution \citep {totanietal08, grauretal11, maozbadenes10, maozetal10, maozetal12, grauretal14}, the absence of hydrogen in the nebular phase \citep {leonard07}, the absence of companions and ex-companions \citep {maozmannuccietal08, gonzalezhernandez12, kerzendorfetal12, schaeferpagnotta12, edwardsetal12, kerzendorf_etal_2014}, and the non-detection of X-ray and radio emission from circumstellar material \citep {lietal11, brownetal12, horeshetal12, bloometal12, marguttietal12, chomiuketal12, marguttietal14, chomiuketal16}. SDs are also found to be inconsistent with observational and theoretical rate predictions \citep {maozetal14}. However, very recent observations have begun to provide strong evidence that near-$M_{\rm ch}$ WD SNe Ia do occur in at least some systems in nature. This evidence has emerged along several fronts. Early light curves in the optical and UV in two events, SN 2012cg and iPTF14atg \citep {caoetal15, marionetal15}, support the interpretation of a shocked companion in these systems \citep {kasen10}.  However, the SD origin of SN 2012cg has been questioned by \citet {shappeeetal16b}.  Some suggestive evidence has also emerged for a possible single-degenerate origin of the Kepler SNR \citep {burkeyetal13, katsudaetal15}. Additional evidence has emerged from the hard X-ray spectra within the supernova remnant 3C 397 \citep{yamaguchietal15}, which is the key focus of this paper.

Evidence  based on the analysis of the hard X-ray spectrum of the galactic supernova remnant (SNR) 3C 397 has provided perhaps the strongest evidence yet in favor of a SN Ia event arising from the SD channel \citep{yamaguchietal15}. A sub-$M_{\rm ch}$ WD progenitor is ruled out for the SNR on the grounds that the observed amount of stable Fe-peak elements cannot be produced by electron capture during C/O burning at densities $\lesssim 10^8$ g cm$^{-3}$, unless the metallicity of the WD progenitor is very high ($> 5.4\ Z_{\odot}$) \citep{yamaguchietal15}.  However, even the spherically-symmetric near-$M_{\rm ch}$ DDT models reported by \citet {yamaguchietal15} also require high WD progenitor metallicity ($> 5\ Z_{\odot}$)  to account for the observed [Mn/Fe] and [Ni/Fe] abundance ratios.

Despite decades of investigation and major progress on both observational and theoretical fronts, many questions about SD SNe Ia remain. What is the internal structure and composition of the progenitor WDs which give rise to SD SNe Ia \citep {umedaetal99, dominguezetal99, dominguezetal01, timmesbrowntruran03, lesaffreetal06, piro08, pirobildsten08}? What does the progenitor WD structure, in particular its central density, imply about the mass accretion history of the WD, and its likely stellar companion \citep {nomoto82a, yoonlanger95, hillmanetal16, starrfieldetal16}? How do near-$M_\text{ch}$ WDs ignite within their convective cores \citep {garciasenzwoosley95, wunschwoosley04, woosleywunschkulhen04,  zingaleetal11, nonakaetal11,  maloneetal14} ? What is the nature of the explosion mechanism in near-$M_\text{ch}$ WDs? Is the detonation mechanism a deflagration-to-detonation transition (DDT) \citep {khokhlov91, hoeflichetal95, niemeyer99, gamezoetal05, ropkeniemeyer07, seitenzahletal13b}, gravitationally-confined detonation (GCD) \citep {plewaetal04, ropkeetal07a, townsleyetal07, jordanetal08, meakinetal09, garciasenzetal16, seitenzahletal16}, a pure deflagration \citep {reineckeetal99, ropkeetal07b}, possibly leaving behind a bound remnant  \citep{jordanetal12b, kromeretal13}, or some other mechanism, such as a pulsational delayed detonation \citep{ivanovaetal74} or a pulsational reverse detonation \citep {garciasenzbravo05, bravogarciasenz06}? What can be learned directly from the observations about the modeling of the complex and still relatively uncertain process of turbulent nuclear combustion \citep {khokhlov95, niemeyerhillebrandt95, niemeyeretal99, lisewskietal00, belletal04a, belletal04b, aspdenetal08, aspdenetal10, aspdenetal11, woosleyetal11, nonakaetal11, zingaleetal11, poludnenkoetal11, maloneetal14}? What do the nucleosynthetic yields of iron-peak elements, including both radioactive species like $^{55}$Fe and stable species like $^{55}$Mn, imply about the overall prevalence of SD SNe Ia, compared to the total SNe Ia rate \citep {seitenzahletal13, yamaguchietal14, seitenzahletal15}? 

Nearly all of these questions have been extensively investigated in the literature, and in particular many theoretical models have been advanced to explain observations of SNe Ia light curves and spectra. However, even when SD models have confronted observations, the comparison has typically been made against Branch normal\footnote{See, e.g. \citet {vaughanetal95} for an early classification of Branch normal SNe Ia based on B-V color at peak, or \citet {branchetal06} for another based on equivalent widths of SiII absorption features at peak.} SNe Ia \citep {hoeflichkhoklov96, ropkeetal12}, which mounting evidence suggests are not typically of SD origin \citep {grahametal15, lundqvistetal15, shappeeetal16}. Consequently,  SD models are not yet tightly constrained by observation. {\it Thus, SNR 3C 397 provides us with an important opportunity to directly confront models of SD SNe Ia against observations of a specific system known to be consistent with a SD origin, and address each of these key questions.}


\citet {yamaguchietal15} explored the physics underlying 3C 397 using 1D DDT models. In this paper, we build upon and extend the modeling of 3C 397 to multidimensional simulations. As previous authors -- e.g. \citet {badenesetal03} -- have noted, multidimensional effects can impact the evolution of a SNR. Most crucially, spherically-symmetric simulations are by necessity only able to consider centrally-ignited ignitions. A body of theoretical work, beginning with a pioneering paper by \citet{garciasenzwoosley95} has suggested that the WD may be ignited off-center, with important ramifications for the ensuing development of the SN Ia \citep {plewaetal04}. A key goal of this current work is to determine whether such off-centered ignitions are consistent with observations of SNR 3C 397.

SD channel models of SNe Ia inhabit a high-dimensional model parameter space. Firstly, the accreting WD progenitor itself may ignite over a range of central densities and compositions, depending upon its mass accretion history \citep {lesaffreetal06}. The progenitor WD metallicity also directly influences the neutron excess during the SN Ia and the resulting abundances of iron-peak isotopes \citep {timmesbrowntruran03}.  Furthermore, turbulent convection within the interior of the WD is an inherently stochastic process,  which may lead to ignition of one or more flame bubbles over a range of offset radii. Authors have variously considered everything from a single ignition point to hundreds or even thousands of ignition points \citep{garciasenzbravo05, livneasidahoflich05, ropkeetal06, ropkeetal07a, ropkeetal07b, kasenropkewoosley09}.  Finally, in addition to the high-dimensionality of the physical  parameter space, one must also consider the systematic uncertainties in the modeling process itself. The systematic uncertainties introduced by the modeling stem from both physical and numerical uncertainties. Examples of physical uncertainties include those associated with the thermonuclear and weak reaction rates \citep {bravoetal11}, and Coulomb corrections to the equation of state \citep {bravogarciasenz99}. Numerical simulations introduce additional modeling uncertainties, including the discrete Eulerian and Lagrangian  \citep {seitenzahletal10} resolutions of the simulation, the geometry and dimensionality of the model, and the subgrid physics assumed for the burning \citep {schmidtetal06}. 

This high-dimensional parameter space of SD models  poses serious challenges -- the ``curse of dimensionality'' -- to attempts  to match SD models against actual SNe Ia events, particularly when one considers 2D and 3D simulations. Firstly, the high-dimensional parameter space of the models makes it very challenging to credibly falsify any specific model, since it is always conceivable that some unexplored corner of parameter space could have produced satisfactory agreement with the observational data. Secondly, even when one is able to obtain a model which matches the observations, degeneracy in the model parameters can make it challenging to conclude that one has obtained a unique match. There may very well be other regions of parameter space, perhaps representing very different physical conditions, which could have also  matched the  observations equally well. For instance, a high progenitor WD metallicity will yield high abundances of iron-peak elements \citep {kruegeretal10, seitenzahletal11, kruegeretal12, seitenzahletal13b}, but so too do more centrally-condensed WDs \citep {woosley97, nomotoetal97, iwamotoetal99}, a greater bubble ignition offset \citep {meakinetal09}, or even carbon-depleted WDs \citep {ohlmannetal14}. Thus, this second issue of model degeneracy poses additional questions as to how we go about definitively connecting realistic multi-dimensional models to observations. 

The approach we adopt in this paper is to employ pragmatic, physically-based constraints to address these challenges, exploring a wide range of possible models. We begin with a standard SD reference model widely considered in the literature, and compare its nucleosynthetic yields against 3C 397. We then systematically explore the physical parameter space, both of the WD progenitor structure and composition, as well as of the ignition. Additionally, we bring in some key physical insights to help reduce the dimensionality of the model space. Recent progress on 3D simulations of ignition in the convective core of near-Chandasekhar mass WDs has revealed that the outcome is typically a single bubble, offset from the center \citep  {zingaleetal11, nonakaetal11,  maloneetal14}. Consequently, using this best-available results from these {\it ab initio} simulations, we constrain our ignitions to be single bubbles, which greatly reduces the model space dimensionality. We do, however, consider a wide range of initial offsets of the flame bubble, from centrally-ignited to the outer edge of the simmering region. Additionally, because the progenitor WD metallicity does not significantly influence the dynamics of SNe Ia \citep {townsleyetal09}, we calculate all hydrodynamical models at zero metallicity, and treat non-zero stellar progenitor metallicity $Z$ during nucleosynthetic post-processing. 

The format of the paper is as follows. In \S \ref {methodology}, we describe the methodology employed to sample the model parameter space of WD progenitors, and to explore a range of possible physical scenarios for the ignition and detonation mechanisms. In section \S \ref {results}, we present the results of our simulations. In \S \ref {discussion} we discuss our findings and conclude.

\section {Methodology}
\label {methodology}

\onecolumngrid
\begin{deluxetable}{l  c c c c c}
\tablecaption{Table of simulation runs presented in this paper.\label{simtable}}
\tablehead{\colhead {Run} & \colhead {Explosion Mechanism} & \colhead {$\rho_c$ (g cm$^{-3}$) } & \colhead {Bubble Offset (km)} &  \colhead {Bubble Radius (km)} & \colhead {C/O Ratio} }
\startdata
DEF-STD & Pure Deflagration & $2.2 \times 10^9$  & 0 & 100 &  50/50 \\ 
GCD-STD & GCD & $2.2 \times 10^9$  & 100 & 16 &  50/50 \\ 
GCD-STD/LOWOFF & GCD & $2.2 \times 10^9$  &  50 & 16  & 50/50 \\
GCD-HIGHDEN & GCD & $6 \times 10^9$  &  100 &  16 & 50/50 \\
GCD-STD/LOWC & GCD & $2.2 \times 10^9$  &  100 & 8 & 30/70 \\
GCD-HIGHDEN-LOWC/HIGHOFF & GCD & $6\times 10^9$  &  200 & 8 &  30/70 \\
DEF-HIGHDEN-LOWC/CENTRAL & Pure Deflagration & $6 \times 10^9$  &  0 & 100 & 30/70 \\
DDT-HIGHDEN-LOWC/CENTRAL & DDT &  $6 \times 10^9$  & 0 &  100 & 30/70 \\  
DDT-HIGHDEN-LOWC/HIGHOFF & DDT &  $6 \times 10^9$  & 200 &  8 & 30/70 \\
\enddata
\explain {Bubble radius column has been added to table.}
\end{deluxetable}

\twocolumngrid

We utilize the Eulerian adaptive mesh refinement (AMR) code FLASH 4.0.1 \citep{Fryxell_2000, dubeyetal09, dubeyetal13}.  We use an equation of state which includes contributions from nuclei, electrons, and blackbody photons, and supports an arbitrary degree of degeneracy and special relativity for the electronic contribution \citep{Timmes_2000}. We include an advection-diffusion-reaction equation treatment of the flame, and incorporate nuclear energy generation using  a simplified treatment of the flame energetics \citep{townsleyetal07, townsleyetal09}. We use a multipole solver \citep {Couch_2013} with isolated boundary conditions and include terms up to $l = 6$ in the multipole expansion for simulations in this paper. To obtain the detailed nucleosynthetic yields from each model, we include Lagrangian tracer particles within our hydrodynamical simulations \citep {dubeyetal12}. The tracer particles are initialized proportional to mass, and passively advected with the fluid. They serve as Lagrangian fluid elements, tracking the hydrodynamic state of the flow throughout the duration of the simulation. All simulations presented here use $10^4$ tracer particles in 2D, roughly equivalent to $10^6$ particles in 3D, a value which has been demonstrated to achieve good precision in previous studies of near-$M_\text{ch}$ WDs \citep {seitenzahletal10}. The Lagrangian tracer particles are subsequently post-processed in the TORCH nuclear network \citep{timmes99}, with 495 species, to obtain the detailed nucleosynthetic yields reported here.  

Our hydrodynamic simulations extend for a few seconds, through the time at which the supernova enters into the free-expansion phase, and further nuclear burning is quenched.  Radioactive isotopes are decayed to the epoch of SNR 3C 397, which we take to be 1750 y \citep {leahyranasinghe16}. \citet {leahyranasinghe16} note the age of the remnant could lie in the range of 1350 - 1750 y, depending upon the uncertain distance to SNR 3C 397, which \citet {leahyranasinghe16} find to be in the range of 8 - 9.8 kpc. However, because no key isotopes have half-lives comparable to the age of the remnant, our nucleosynthetic yields are insensitive to the age of the remnant.  For instance, in the decay chain $_{27}^{55}\textrm{Co}\;\overset{\mathrm{18 h}}\longrightarrow\;_{26}^{55}\textrm{Fe}\;\overset{\mathrm{3 y}}\longrightarrow\;_{25}^{55}\textrm{Mn}$, $^{55}_{27}$Co is effectively fully decayed to $^{55}_{25}$Mn. In contrast, for the decay chain $_{26}^{53}\textrm{Fe}\;\overset{\mathrm{9 m}}\longrightarrow\;_{25}^{53}\textrm{Mn}\;\overset{\mathrm{4 My}}\longrightarrow\;_{24}^{53}\textrm{Cr}$, we decay $_{26}^{53}\textrm{Fe}$ only to $_{25}^{53}\textrm{Mn}$ \citep {unterweger92}.

 
Recent studies have compared individual 3D numerical simulations against observations \citep {seitenzahletal16}. Such 3D simulations have numerous advantages over 2D simulations, since they generally capture a greater degree of realism, including, for instance, enhanced flame surface area and burning \citep {ropkeetal07b, jordanetal08}. 3D simulations also faithfully capture the physics of the turbulent energy cascade, which is inverted in the case of 2D turbulence \citep {kraichnan67}. However, such realism in full 3D comes with a trade-off, since each model is much more computationally expensive than 2D models, and as a consequence  results in a reduced ability to explore the model parameter space. In the present study, we employ 2D axisymmetric hydrodynamical models to enable a greater exploration of the model parameter space. As we will see, because the nucleosynthetic yields of the iron-peak elements depends sensitively upon the progenitor WD structure, the ignition condition, and the detonation mechanism, such an exploration of model parameter space is critical in confronting simulations with SNR 3C 397. 

We fix the initial WD C/O abundances at the start of the hydrodynamic evolution. The effect of the WD progenitor metallicity is modeled in nucleosynthetic post-processing using the Torch code \citep {timmes99} by the addition of $^{22}$Ne, which serves as a replacement for metallicity-dependent neutron excess \citep {timmesetal03, milesetal16}. All abundances are scaled to solar using \citet {asplundetal09}. 
The initial convective phase leading up to ignition is also expected to produce neutron enrichment. We do not model this convective neutronization;   it is, however, expected to be significant  for $Z < Z_{\odot}/ 3$ \citep {martinezrodriguezetal16}. \added {Additionally, by fixing the initial abundances during post-processing, we are in effect neglecting electron captures from the small sparks at the true onset of ignition within the WD, up until the much larger flame bubbles which we must adopt by necessity in our hydrodynamic simulations. The size distribution of these ignition sparks remain largely uncertain, and have been estimated to be anywhere between 10 cm - 1 km in spatial extent \citep {woosleywunschkulhen04}. However, by repeating DDT-HIGHDEN-LOWC/CENTRAL with a smaller but still-resolved bubble 16 km in radius, we have determined these missed electron captures amount to less than one part in $10^3$ in the mass-weighted mean electron fraction of the WD. Because the mass interior to the bubble scales as the bubble radius cubed, smaller initial bubble radii than 16 km have a negligible impact on this finding, so that one part in $10^3$ is an upper-bound on the mean mass-weighted electron abundance error introduced by the missed electron captures. Further, In our offset models, the flame bubble contains a very small amount of mass initially, so the missed electron capture effect is orders of magnitude less in these instances.}

We further note that ONe WDs have high central densities comparable to some of our models for 3C 397, but ignite at central densities exceeding $\rho_c \sim 10^{10}$ g cm$^{-3}$\citep {schwabetal15}. The fate of such electron capture supernovae in ONe WD progenitors has only recently begun to be explored in 3D simulations, and suggest a range of outcomes including complete disruption, stable bound remnants, and accretion-induced collapses to neutron stars. These current models do not resemble SNe Ia  \citep {jonesetal16}. Thus we exclude consideration of ONe WDs as progenitors for 3C 397 in this paper.

Models similar to GCD-STD have been studied widely in the literature. Assuming that a deflagration-to-detonation transition is not triggered during the buoyant eruption of the bubble from the surface of the WD, these models are found to lead to a gravitationally-confined detonation \citep {plewaetal04, townsleyetal07, jordanetal08, meakinetal09, seitenzahletal16}.  We build upon the standard model GCD-STD, and systematically and individually  vary the most significant parameters which determine the progenitor WD structure, the ignition, and the detonation mechanism itself. These parameters include the progenitor WD central density $\rho_c$ and its carbon/oxygen ratio, as well as the ignition offset -- see Table \ref {simtable}.  Specifically, we consider variants of the standard model GCD-STD with a lower ignition offset (GCD-STD/LOWOFF), a higher central density (GCD-HIGHDEN), and a lower carbon/oxygen fraction (GCD-STD/LOWC). We further consider a model variant including the combined effects of lower ignition offset, higher central density and lower carbon/oxygen fraction (GCD-HIGHDEN-LOWC/HIGHOFF). We also consider differing explosion mechanisms, including two pure deflagration models (DEF-STD and DEF-HIGHDEN-LOWC/CENTRAL) as well as two deflagration-to-detonation transition models (DDT-HIGHDEN-LOWC/CENTRAL and DDT-HIGHDEN-LOWC/HIGHOFF). For our DDT models, we fix the deflagration-to-detonation transition density to be $2.6 \times 10^7$ g cm$^{-3}$. Our DDT model setup is described in \citet {townsleyetal09}, \citet{kruegeretal10}, and \citet {jacksonetal10}. 


Our 2D $r-z$ domain assumes azimuthal symmetry about the $z$ axis,  extends from $-6.5536\times 10^{5}$ km to $+6.5536\times 10^{5}$ km in the $z$ direction, and ranges from 0 to $+6.5536\times 10^{5}$ km in the $r$ direction. \added {The finest linear spatial resolution in all models is 4 km.} A very low density region surrounding the WD, sometimes referred to in the literature as ``fluff,'' is required by Eulerian grid-based simulations, which cannot treat empty space without some matter density.  The fluff is chosen to have an initial density of $10^{-3}$ g cm$^{-3}$, a temperature of $3 \times 10^7$ K, and is dynamically unimportant for the duration of the models presented here.  

We employ several refinement criteria following \citet {townsleyetal07} and \citet {townsleyetal09}, which are designed to follow the nuclear burning of the models at high resolution, while also minimizing the resolution in the very low density regions outside the WD itself. Our simulations seek to maintain the highest resolution in the burning region behind the flame surface, and employ a standard dimensionless density gradient criterion to refine when the dimensionless density gradient parameter exceeds 0.1, and derefines when it is beneath 0.0375. Further refinement criteria seek to derefine in the fluff and in regions outside of active burning, derefining one level if the energy generation rate is lower than $5 \times 10^{17}$ erg g$^{-1}$ cm$^{-3}$, and completely to the base level if the density is below $10^3$ g cm$^{-3}$. 

A well-known artifact when coupling a real stellar EOS with hydrodynamics is the development of temperature oscillations in the vicinity of discontinuities \citep {zingalekatz15}. In the current context, these temperature oscillations can lead to spurious burning and even artificial detonations. The effect is most pronounced  once the flamelet becomes Rayleigh-Taylor unstable and enters the turbulent phase of burning.  In order to avoid spurious burning resulting from such temperature oscillations, we only allow burning outside the flame in a cone which opens up at the south pole, opposite of the breakout, with a half-opening angle of $\sim 20^\circ$. Once the hot ash sweeps across the surface of the WD and enters this cone, this burning suppression condition is relaxed.

\section {Results}
\label {results}


\subsection {Neutronization and Electron Abundance $Y_e$}

\begin{figure*}[h]
	\begin{center}
		\includegraphics[width=1.5\columnwidth]{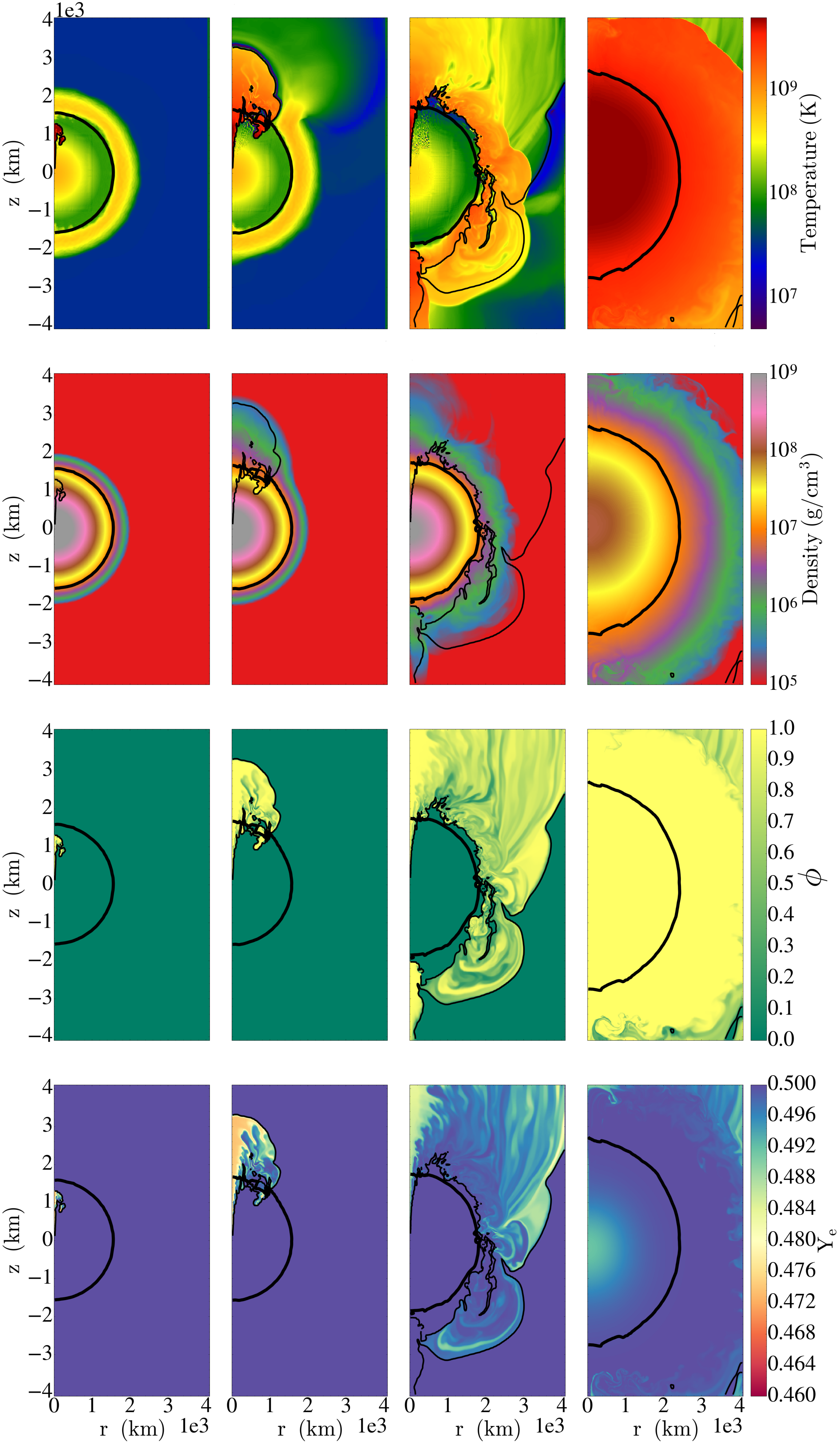}
		\caption{Plots of the temperature, density, flame fraction of burned material $\phi$, and  electron fraction $Y_\mathrm{e}$, respectively, from top to bottom, for the GCD-STD run. From left to right, these quantities are depicted  at times $t = 0.7$ s, 1.0 s, 2.0 s, and 2.5 s, respectively. The thick black line is an isocontour of $10^7$ g/cm$^3$ in density and the thin black line is an isocontour of $0.1$ in the burned fraction $\phi$ approximately depicting the density at which detonation arises, and the fuel-ash boundary, respectively. Note that the thin black line demarcating the fuel-ash boundary lies partially out-of-frame subsequent to bubble breakout at times $t = 2.0$ s and 2.5 s due to the ejection of ash from the WD.}
		\label {fig:gcd_std}
	\end{center}
\end{figure*}

Figure \ref {fig:gcd_std} depicts the evolution of the GCD-STD run, viewed through four key physical fields, at four key evolutionary times $t =$0.7 s, 1.0 s, 2.0 s, and 2.5 s. The fields are the temperature, density, the burned fraction $\phi$, a scalar field tracking the flame surface which ranges between 0 for pure fuel and 1 for pure ash \citep {townsleyetal16}, and the electron abundance is defined as the weighted sum over isotopes $i$ of the ratio of atomic number $Z_i$ to atomic mass number $A_i$, $Y_e = \sum_i X_i Z_i / A_i$, where $X_i$ is the isotopic mass abundance. The first column in Figure \ref {fig:gcd_std} at $t = 0.7$ s shows these fields as the 100 km offset bubble rises buoyantly and approaches  breakout on the surface of the WD. In the second column at $t = 1.0$ s, the hot deflagration ash expands once it breaks out of the WD and  sweeps across the surface; meanwhile, the WD pre-expands due to the energy released by this burning. In the third column at $t = 2.0$ s, the hot ash converges at the point opposite of breakout at the onset of the detonation. Lastly, in the fourth column, the detonation wave has completely swept through the WD at $t = 2.5$ s.

The $Y_e$ panels at the bottom of Figure \ref {fig:gcd_std} reveals that neutronization ($Y_e < 0.5$) occurs in two principal stages. Firstly, as seen in the first panel panel at $t = 0.7$ s, during the initial deflagration phase, neutronization occurs within the flame bubble. Later, in the second and third panels at $t = 1.0$ s and $t = 2.0$ s, the bubble is subsequently buoyantly expelled from the WD interior, leading to the detonation. In the fourth panel at $t = 2.5$ s, depicting the post-detonation phase, the centermost region of the WD also undergoes significant neutronization during the passage of the detonation front. While these results were calculated specifically for the GCD-STD model, the underlying physics is similar for other models as well. In particular, in agreement with previous work, we confirm that a SN Ia undergoing a DDT also experiences significant neutronization in both the deflagration and detonation phases along broadly similar lines \citep {seitenzahletal11, seitenzahletal13b}. In particular, a single buoyancy-driven ignition point generally leads to a  low-deflagration SD SN Ia {\it in both the DDT and GCD scenarios}. In such ignitions, deflagration minimally pre-expands the density profile of the progenitor, and thereby primes the WD for a substantial iron group element (IGE) yield during the subsequent detonation phase, {\it largely irrespective of the detonation mechanism itself.} However, subtle distinctions between the GCD and DDT models do exist.

From the standpoint of the production of neutron-rich IGE, the key distinction between the GCD and the DDT detonation mechanisms is that a DDT is posited to transition to a detonation {\it prior} to bubble breakout, whereas a GCD undergoes a detonation only {\it subsequent} to bubble breakout and ash wraparound.  Consequently, for a fixed ignition in a given WD progenitor,  a DDT will {\it always} undergo less pre-expansion prior to detonation, and will {\it always} have a higher central density at detonation than a GCD. \footnote {A WD undergoing a pulsationally-assisted  GCD mechanism of \citet {jordanetal12a} does experience a significant re-compression subsequent to bubble ejection, but requires a significantly higher deflagration energy release with multiple ignition points than considered here.} This key difference between the DDT and GCD mechanisms in turn implies that for identical deflagration phases leading up to detonation, a DDT will generally produce greater neutronization than a GCD.  

\begin{figure*}[h]
\begin{center}
\includegraphics[width=2.0\columnwidth]{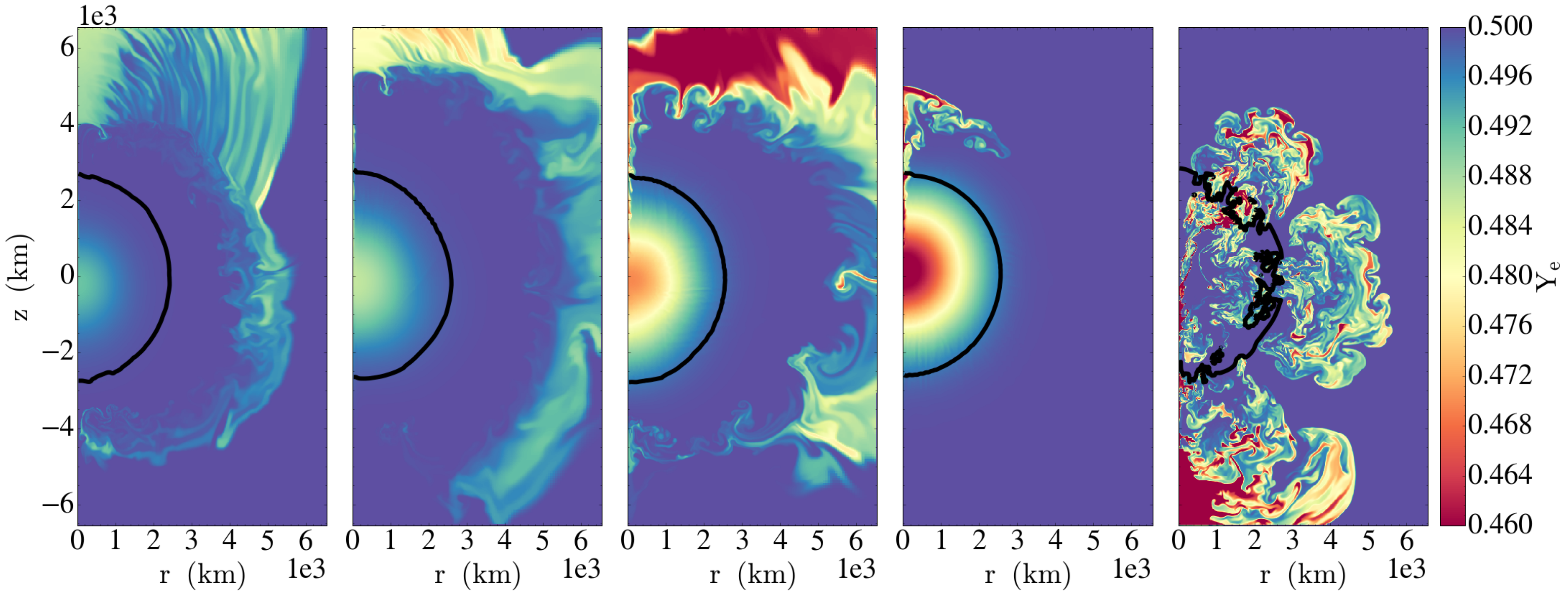}
\caption{Slice plots of $Y_\mathrm{e}$ for representative models after the maximum density in each model has dropped below $10^8$ g cm$^{-3}$. The models, from left to right, are GCD-STD, GCD-STD/LOWC, GCD-HIGHDEN-LOWC/HIGHOFF, DDT-HIGHDEN-LOWC/HIGHOFF, and  DEF-HIGHDEN-LOWC/CENTRAL, respectively. The thick black line is an isocontour of $10^7$ g/cm$^{-3}$ in density, as in Figure \ref {fig:gcd_std}.}
\label {fig:ye}
\end{center}
\end{figure*}

This distinction between the various explosion models is illustrated in Figure \ref {fig:ye}, where representative plots of the electron fraction $Y_e$ are shown for five different models: three GCDs, a DDT, and a pure deflagration. The plots are all taken at a comparable stage of evolution, when the central density of the white dwarf has dropped below $10^8$ g cm$^{-3}$. It is evident that lowering the carbon-oxygen fraction in GCD-STD/LOWC and increasing the offset of the ignition bubble in GCD-HIGHDEN/LOWC enhances the neutronization over the baseline model GCD-STD, as we discuss further in \S 3.4 and \S 3.5. Similarly, it can also be seen that the ejected IGEs produced within the GCD models will be ejected at larger velocities than the model DDT-HIGHDEN-LOWC/HIGHOFF in the fourth panel of Figure \ref {fig:ye}. The last panel of Figure \ref {fig:ye} shows the strongly asymmetric distribution of neutronization in the pure deflagration model DEF-HIGHDEN-LOWC/CENTRAL.  

\subsection {Stable Iron Group Element Production}

\begin{figure*}[h]
	\begin{center}
		\includegraphics[width=2.0\columnwidth]{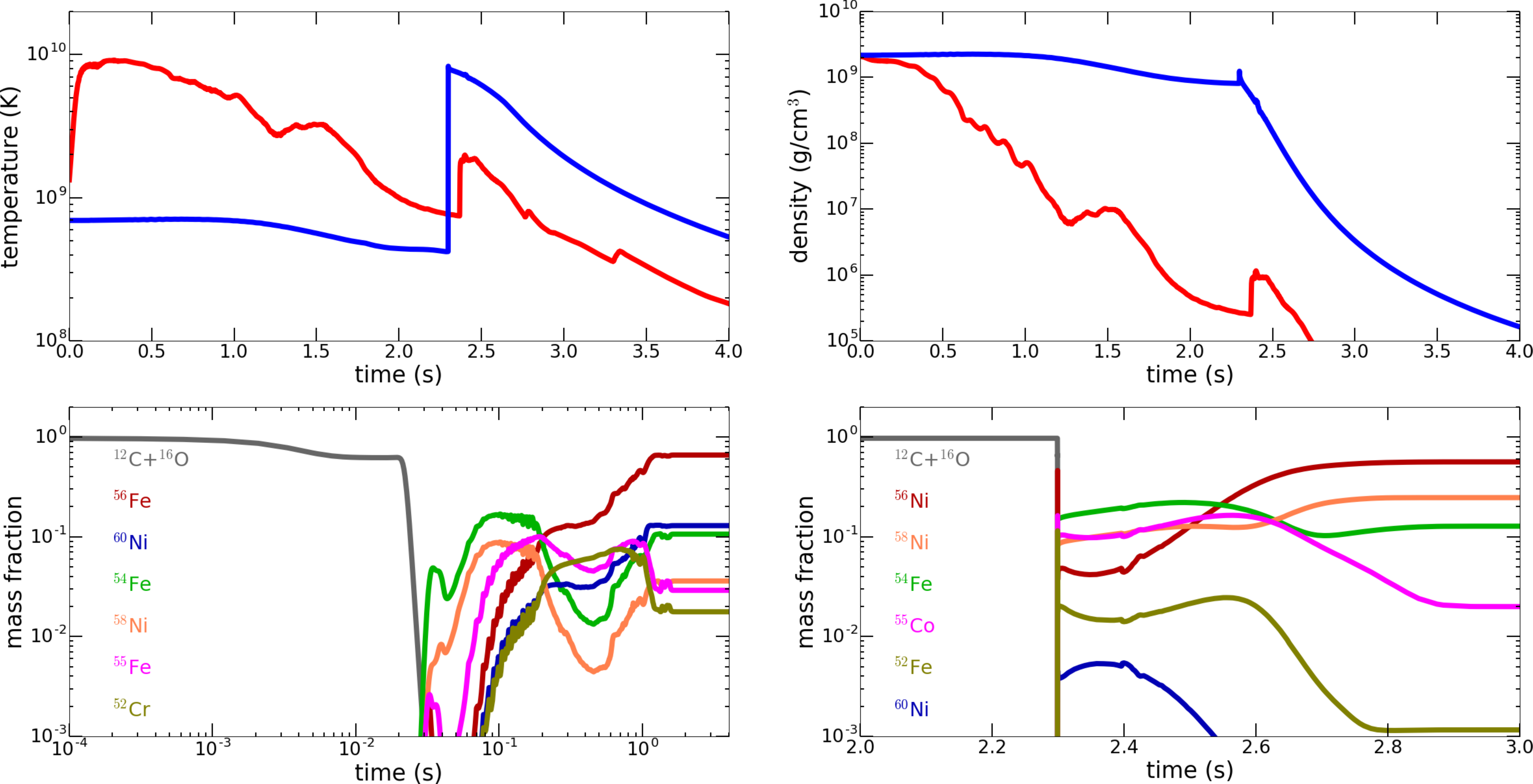}
		\caption{Sample Lagrangian tracer particle trajectories, depicting the thermodynamic and nucleosynthetic evolution of two individual fluid elements -- one undergoing a deflagration and then shocked by the detonation shock (red),  and another only a detonation (blue). The lower-left panel depicts the red particle nucleosynthetic abundances over time, and the lower-right panel shows the blue particle nucleosynthetic abundances over time.}
		\label {fig:sample_lagrangians}
	\end{center}
\end{figure*}

The direct production of stable iron group elements occurs during burning at high densities within the normal freeze-out regime, which is characterized by low entropy burning \citep {thielemannetal86}. We now illustrate the production of stable iron group elements in both the deflagration and the detonation phases using two sample Lagrangian trajectories from the GCD-STD model.  
Figure \ref {fig:sample_lagrangians} depicts the thermodynamic history and nucleosynthetic production of iron group elements within two fluid elements. The particle shown in the red hydrodynamic curves on the top panels and on the bottom left nucleosynthetic panel  of Figure \ref {fig:sample_lagrangians} is swept up by the flame bubble by $t = 0.01$ s, and then later at $t = 2.3$ s, it encounters the detonation shock. For comparison, a particle shown in the blue hydrodynamic curves and on the bottom right nucleosynthetic panel of Figure \ref {fig:sample_lagrangians} never encounters the deflagration front, but is also detonated just prior to the red particle. 

The top two panels of Figure \ref {fig:sample_lagrangians} depict the hydrodynamic evolution of the Lagrangian particles in both temperature (left) and density (right). Both particles have nearly the same initial density, $2 \times 10^9$ g/cm$^3$, and are both initially located within the central 120 km of the WD. The red particle's temperature rapidly climbs as it is deflagrated. By $t = 0.2$ s, the deflagration has consumed all  C/O fuel, and produced mass fractions of $\sim 0.1$ for $^{54}$Fe, $^{56}$Fe, and $^{55}$Fe, characteristic signatures of neutronization during burning in the low-entropy normal freezeout regime \citep {thielemannetal86}. Significant abundances of $^{52}$Cr, $^{60}$Ni, and $^{55}$Fe are also produced during deflagration. Subsequent to deflagration, but prior to detonation, from $t = 0.2$ s to $t = 1.0$, the abundances continue to evolve under the influence of weak interactions and adiabatic expansion of the hot bubble material under shifting NSE conditions \citep {calderetal07}. The parcel of fluid represented by the particle is expelled from the WD on the northern hemisphere of the WD.  Just prior to the passage of the detonation front over the parcel at $t = 2.4$ s, its density has reached $2 \times 10^5$ g/cm$^3$, its temperature is  $7 \times 10^8$ K, and the compositions have frozen out. The passage of the detonation shock over the burned material does not significantly alter the composition. The composition has shifted from the initial deflagration to be dominated by $^{56}$Fe, about 0.55 by mass fraction. The next most-abundant species are $^{60}$Ni, $^{54}$Fe,  $^{58}$Ni, $^{55}$Fe, and $^{52}$Cr -- all stable isotopes with the exception of $^{55}$Fe, which decays to $^{55}$Mn. The resulting decayed IGE abundance ratios for this deflagrating and detonating particle are Ni/Fe $= 0.21$, Mn/Fe $ = 0.037$, and Cr/Fe $ = 0.029$, all roughly in the same range as observed in SNR 3C 397. 

The blue curves on the top panels and on the bottom right nucleosynthetic panel of Figure \ref {fig:sample_lagrangians} depict a particle which undergoes a detonation only. While the initial density of this particle is comparable to that of the red particle, both $\sim 2 \times 10^9$ g/cm$^3$, the blue particle is outside the initial flame bubble, and has a lower initial temperature of $7 \times 10^8$ K. From $t = 0$ s to $t = 2.3$ s, the blue particle undergoes an adiabatic expansion as the WD pre-expands during the deflagration phase. Unlike the red particle, which is ejected from the core of the WD, the blue particle remains near the center of the WD throughout. The detonation front, which originates near the south pole of the WD in this case, passes over the blue particle at a slightly earlier time, with the time difference corresponding roughly to the detonation-crossing time of the pre-expanded WD, $R_{\rm WD} / v_{\rm CJ} \sim 0.1$ s, where $R_{\rm WD} \sim 2 \times 10^3$ km, and $v_{\rm CJ}$ is the Chapman-Jouguet  velocity $v_{\rm CJ} \sim 1.6 \times 10^4$ km/s. The composition of the blue particle remains pure fuel until the passage of the detonation front at $t = 2.3$ s. The dominant product of the detonation burning is $^{56}$Ni, with a mass fraction $\sim 0.55$, followed by the less abundant products $^{58}$Ni, $^{54}$Fe,  $^{55}$Co, $^{52}$Fe, and a trace level of $^{60}$Ni. Both $^{55}$Co and $^{52}$Fe are radioactive; $^{55}$Co decaying first to $^{55}$Fe and thence to $^{55}$Mn, and $^{52}$Fe decaying to $^{52}$Mn, followed by a decay to $^{52}$Cr. The detonation of this parcel consequently yields decayed stable IGE abundance ratios of Ni/Fe $=0.34$, Mn/Fe $= 0.027$, Cr/Fe $=0.0025$. Notably, while the Ni/Fe and Mn/Fe ratios  for this trajectory are roughly in the same range as observed in SNR, the Cr/Fe ratio is lower by an order of magnitude.

\begin{figure*}[h]
	\begin{center}
		\includegraphics[width=2.0\columnwidth]{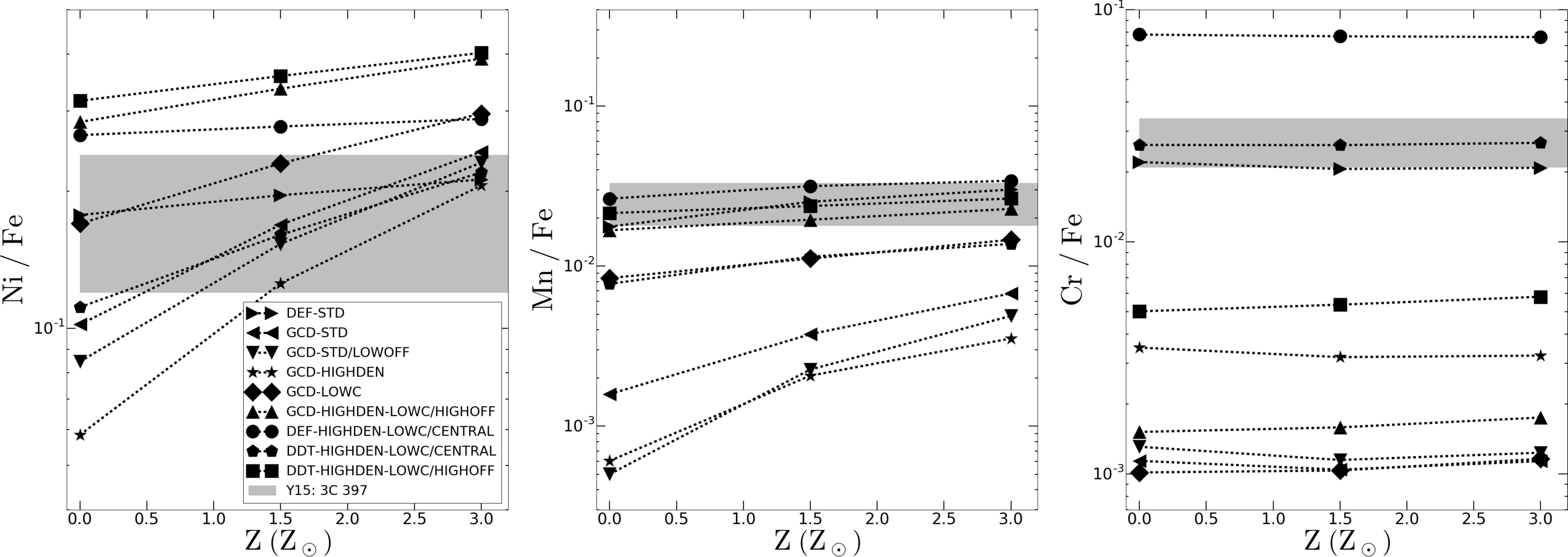}
		\caption{Comparison of computed stable-iron peak abundance ratios for Ni/Fe, Mn/Fe, and Cr/Fe versus those observed for SNR 3C 397. The grey bands on each figure indicate the measured value of each abundance ratio, within one $\sigma$ observational error bars, as reported in \citet {yamaguchietal15}. Figure symbols are indicated in the legend.}
		
		\label {fig:ige_ratios}
	\end{center}
\end{figure*}

\begin{figure*}[h]
	\begin{center}
		\includegraphics[width=2.0\columnwidth]{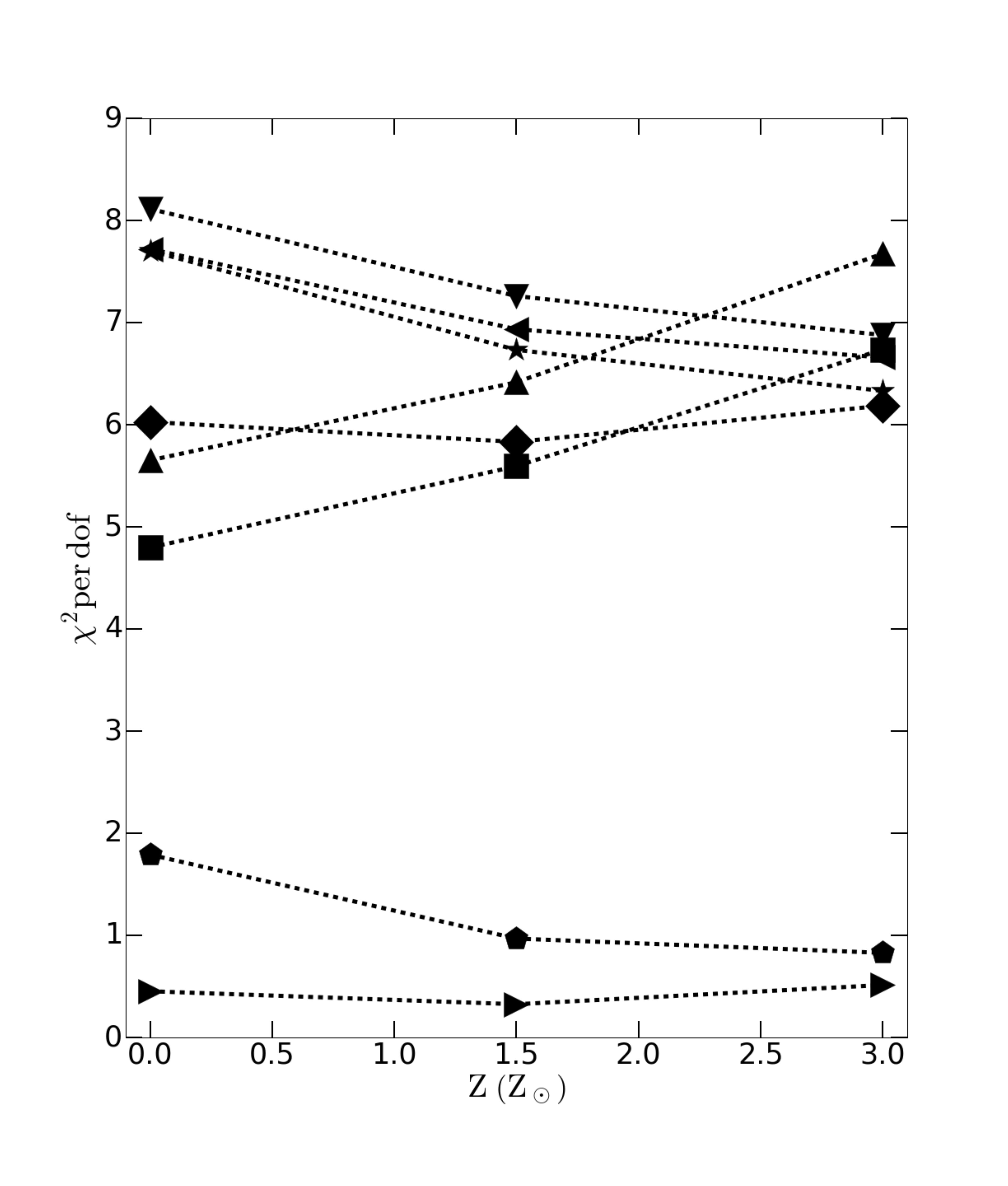}
		\caption{Aggregate $\chi^2$ error of model nucleosynthetic yields compared to SNR 3C 397, for all models except DEF-HIGHDEN-LOWC/CENTRAL, whose errors exceed these values by over an order of magnitude. Figure labels are the same as figure \ref {fig:ige_ratios}.}
		\label {fig:chi2}
	\end{center}
\end{figure*}


\begin{figure*}[h]
	\begin{center}
		\includegraphics[width=2.0\columnwidth]{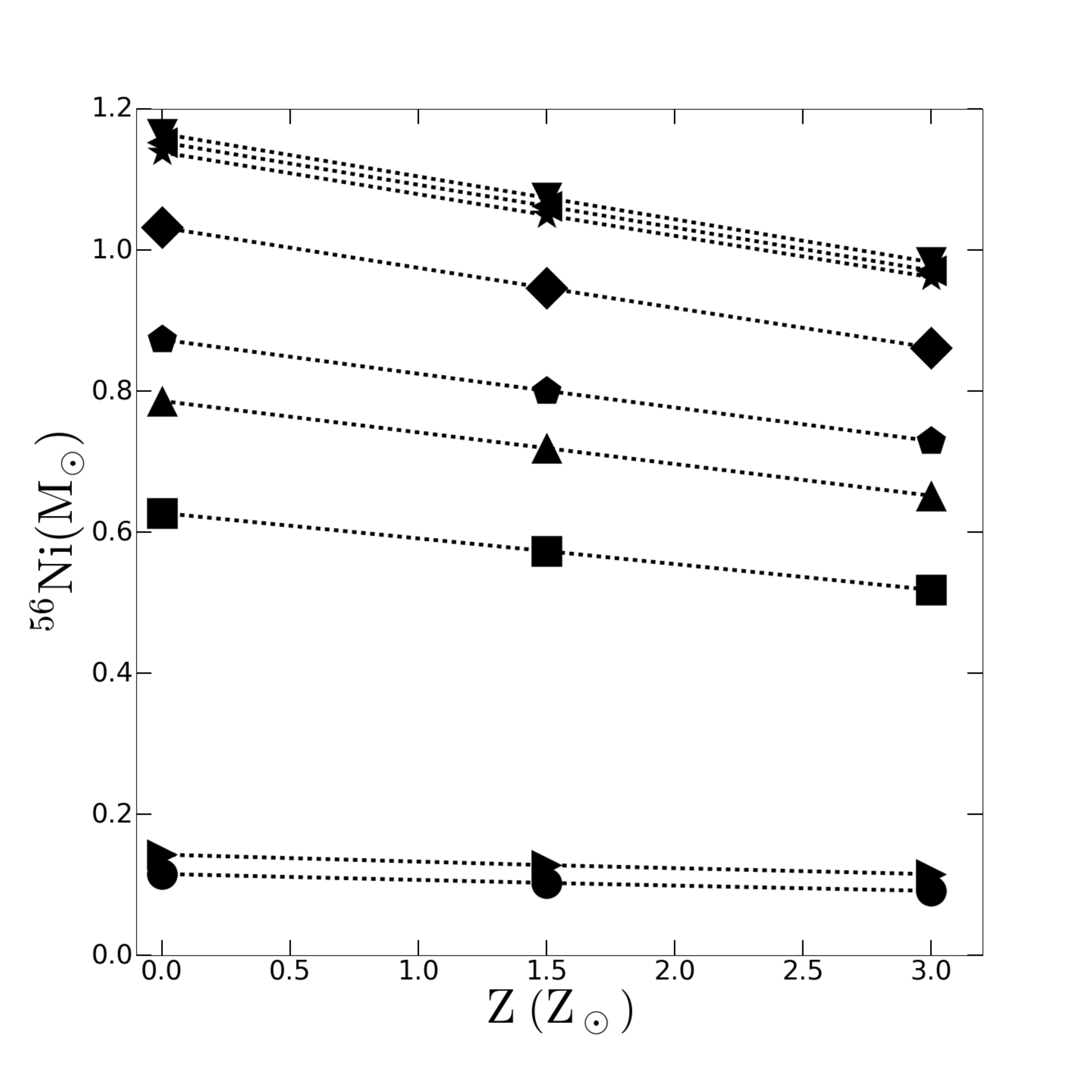}
		\caption{Mass of $^{56}$Ni synthesized for each model and metallicity. Figure labels are the same as figure 5.}
		\label {fig:ni56}
	\end{center}
\end{figure*}

\begin{figure*}[h]
	\begin{center}
		\includegraphics[width=2.0\columnwidth]{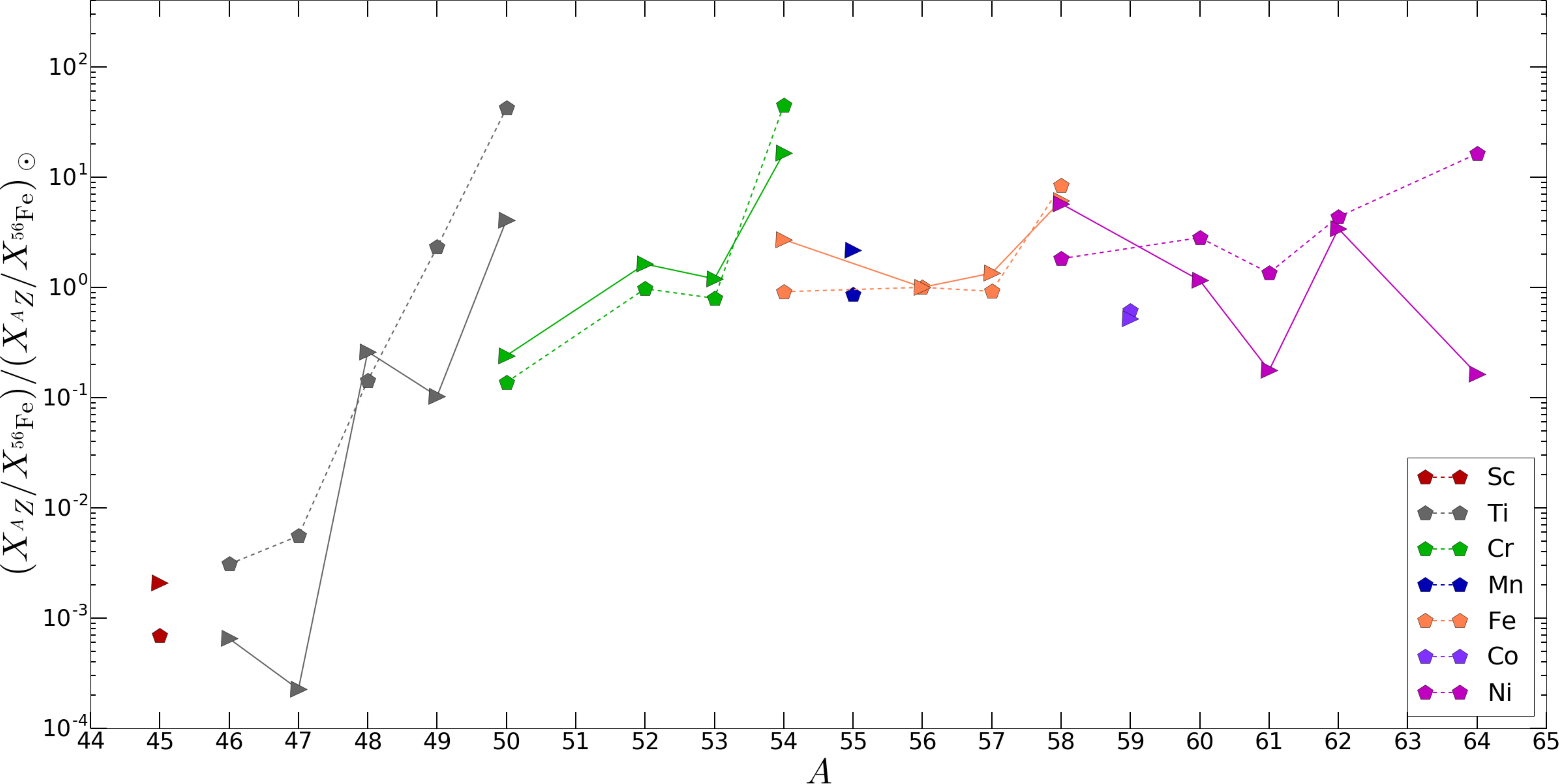}
		\caption{Model remnant isotopic abundance ratios of IGE abundances  to $^{56}$Fe, relative to solar, as a function of nucleon mass number $A$. The models DEF-STD (right triangles) and DDT-HIGHDEN-LOWC/CENTRAL (pentagons) are plotted. The plot points are colored by chemical species, as indicated in the legend.}
		\label {fig:isotopic_abundances}
	\end{center}
\end{figure*}

The overall level of agreement of the numerical simulations with the observations of SNR 3C 397 may be quantified by the squared summed errors in each of the three IGE ratios, normalized to the experimental standard deviation, and divided by the three degrees of freedom provided by the three abundance ratios. That is, we define the aggregated squared summed error $\chi^2$ per degree of freedom between the observational and modeled results as
\begin {equation}
\chi^2 = {1 \over 3}  \sum_{i = 1}^3 \left ({X_i - X_{i, \rm 3C 397} \over \sigma_i} \right)^2
\end {equation}
Here $X_i$ is the model abundance ratio, $X_{i, \rm 3C 397}$ is the observed abundance ratio for SNR 3C 397, and the index $i$ runs from 1 to 3, tracking each of the abundance ratios for Ni/Fe, Mn/Fe, and Cr/Fe. The $\chi^2$ metric defined in this way incorporates only the observational error bars as determined by \citet {yamaguchietal15}, and does not include the systematic errors associated with the numerical models.  
We discuss the results for $\chi^2$ in the sub-sections \S 3.3 - 3.7.

\subsection {Effect of Varying Ignition Offset}

We turn our attention to the question of how near-$M_{\rm Ch}$ WDs ignite within their convective cores. As we have seen, both deflagration and detonation produce stable IGEs at high densities in near-$M_{\rm ch}$ WDs in the normal freezeout regime, and the alpha-rich NSE regime \citep {thielemannetal86}. Yet there is an essential competition between deflagration and detonation, because a greater deflagration energy release yields a greater pre-expansion of the progenitor WD. The greater pre-expansion in turn leads to lower central densities during the subsequent detonation phase, and hence a lower yield of stable IGEs for the detonation. 

We note that ignition within the turbulent convective core of the WD interior is inherently stochastic, and is expected to lead to a range of ignition offsets. We examine the role of deflagration by  comparing the IGE yields of two models, varying the ignition offset radius -- the GCD-STD and the GCD-STD/LOWOFF models. Specifically, we have chosen the GCD-STD model to have an ignition offset of 100 km, and  the model variant GCD-STD/LOWOFF to have a lower ignition offset radius of 50 km -- see Table 1. Multidimensional direct numerical simulations -- see e.g. \citet {zingaleetal11, nonakaetal11,  maloneetal14} -- suggest a range of ignition offsets from 0 to 100 km for this central density, with a mean value of 50 km, and a likely range of 40 to 75 km. However, to date these have been the only large-scale studies, on a single WD progenitor with a fixed composition, of the crucial problem of turbulent convection in near-$M_{\rm Ch}$ WD interiors. Consequently, for the purposes of the current validation study,  we span the range of offsets predicted by the ignition simulations, and extend beyond it as well.

The stable IGE abundance ratios Mn/Fe, Ni/Fe, and Cr/Fe are shown in Figure \ref {fig:ige_ratios} as a function of metallicity $Z$ relative to solar. Strikingly, we find that both models GCD-STD and GCD-STD/LOWOFF significantly underproduce all three abundance ratios, by factors of up to two orders of magnitude. The enhanced neutron excess for the higher metallicities considered ($Z = 1.5, 3 Z_{\odot}$) yield somewhat better agreement for Ni/Fe, but are significantly off for Mn/Fe and Cr/Fe. $^{52}$Cr, the predominant Cr isotope, is produced by the radioactive decay of the parent nucleus $^{52}$Fe. $^{52}$Fe is  connected in NSE to $^{56}$Ni through the  reaction $^{52}$Fe$(\alpha, \gamma)$
$^{56}$Ni \citep {badenesetal08}. Consequently, the Cr/Fe production is nearly independent of metallicity $Z$, and enhancing metallicity does not improve the agreement of the Cr/Fe ratio. Moreover, GCD-STD/LOWOFF, whose ignition offset radius is considered to be more likely on the basis of numerical simulations -- see e.g. \citep {nonakaetal11} -- underproduces all stable IGE abundance ratios relative to 3C 397 even more than GCD-STD at all metallicities considered. 

The resulting $\chi^2$ per degree of freedom is plotted in Figure \ref {fig:chi2} as a function of metallicity $Z$ for all models considered in this paper except for DEF-HIGHDEN-LOWC/CENTRAL, whose errors are larger than the models considered here, and would fall well outside the plot. 
The model GCD-STD/LOWOFF, shown in down triangles, has greater aggregated deviation from the observations, as quantified by $\chi^2$, across all metallicities. 

These results may be understood by considering that a single ignition bubble with an ignition offset exceeding 20 km is buoyancy-driven, and typically burns only a small fraction of the mass of the star during the deflagration phase \citep {fisherjumper15}. In a nutshell, less expansion leads to greater IGE production. Thus, while some stable IGEs may be produced during deflagration of a buoyancy-driven bubble, a significant amount of fuel is left behind. In contrast, it is the subsequent detonation phase, which encompasses the entire WD, which can potentially yield much greater stable IGE abundances. Hence,  a low deflagration yield, with low pre-expansion, is necessary in order to achieve the highest possible stable IGE abundances for a given WD progenitor. This is essentially why a larger ignition offset radius leads to larger stable IGE abundance ratios.

\subsection {Effect of Lowering the WD Carbon/Oxygen Ratio}

Next, we address the question of the internal composition of the progenitor WD, by examining the influence of the WD progenitor C/O fraction on the production yields of stable IGEs. In the literature, progenitor model WDs for SD SNe Ia are often assumed to have equal abundances of carbon and oxygen,
 even though stellar evolution calculations suggest the C/O fraction of WDs should be lower, due both to normal stellar evolution \citep {umedaetal99, dominguezetal99, dominguezetal01} and the subsequent simmering phase as the WD approaches $M_{\rm ch}$  \citep{lesaffreetal06}. The C/O fractions predicted by stellar evolutionary models remain relatively uncertain, because of the uncertainty in the \deleted {in the} nuclear reaction rate of $^{12}$C $(\alpha, \gamma) ^{16}$O \citep {fieldsetal16}, and the uncertainty in the modeling of turbulent convection, which may impact the size of the convective core during the core helium burning phase \citep {dominguezetal01}. 

The C/O fraction directly impacts SD SNe Ia models due to the influence upon the laminar flame speed. Lower C/O fractions yield a lower laminar flame speed \citep {timmeswoosley92}, and hence a less vigorous deflagration. Perhaps even more significantly, the lower C/O fraction also decreases the energy release in the flame front and therefore reduces the buoyancy and the amount of expansion during deflagration. As shown in \citet {wilcoxetal16}, a lower C/O ratio should also lead to a lower overall IGE yield, mainly through the impact upon the lower-density portion of the burn. These effects lead to higher mass fractions of stable IGE. Here we make the simplifying assumption that the C/O fraction is uniform throughout the WD interior. 

We compare the stable IGE abundance ratios for our standard GCD-STD against a second model, GCD-STD/LOWC identical in all respects to GCD-STD except for a lower C/O fraction of 30/70. The abundance ratios for these models are plotted as a function of metallicity in  Figure \ref {fig:ige_ratios}, with GCD-STD in left-triangles, and GCD-STD-LOWC as diamonds. The lower C/O fraction for model GCD-STD/LOWC yields greater IGE abundances, and stable IGE abundance ratios which are closer to the SNR 397 than GCD-STD across all metallicities, as seen in Figure \ref {fig:chi2}. 

\subsection {Effect of Increasing WD Central Density}

We now delve into the issue of the progenitor WD structure. The central density of near-$M_{\rm ch}$ WDs at ignition is a relatively uncertain parameter that enters into SD scenario simulations. \citet {lesaffreetal06} determined the central density at ignition for a wide range of WD models under the assumption that the mass accretion rate onto the WD is regulated by a Hachisu wind.  In their work, \citet {lesaffreetal06} demonstrate that a broad range of central densities at ignition of $2 - 5 \times 10^9$ g cm$^{-3}$, are produced depending on the initial WD mass.  Because the electron capture rates are highly sensitive to density, an increase of the WD progenitor central density should in principle enable greater production of stable IGEs in the normal freezeout regime. 

We compare the stable IGE abundance ratios for our standard GCD-STD against GCD-HIGHDEN. Model GCD-HIGHDEN varies only the initial central density from GCD-STD, setting it to $\rho_c = 6 \times 10^9$ g/cm$^3$, and a total mass of $1.3987 M_{\odot}$. We plot the abundance ratios of GCD-HIGHDEN in stars in Figure \ref {fig:ige_ratios}. It is apparent that despite the enhancement of the initial central density of the progenitor WD, and the benefit of the strong density dependence of the electron capture reactions, the abundance ratios are in all cases strongly suppressed relative to GCD-STD. A comparison of the $\chi^2$ per degree of freedom in Figure \ref {fig:chi2} shows that the agreement for GCD-HIGHDEN is the poorest among all detonating model variants considered here, across all metallicities. 

The explanation for this seemingly counterintuitive decrease in abundance ratios with an increase in central density stems from the fact that even as the density boosts the electron capture rate, so too does it enhance the deflagration energy release. Specifically, the central density enhances the laminar flame speed, and hence the deflagration energy release and the pre-expansion experience by the WD. Hence, in order to achieve this density enhancement of stable IGEs, the WD must simultaneously suppress the natural tendency of the carbon flame speed to increase with increasing density. To reduce deflagration energy release, we must look towards the chemical composition of the WD, and to the ignition.

 \subsection {Combined Effects of Low C/O, High Central Density, High Ignition Offset}
 
We build upon the results described in previous sections for a single WD progenitor model variant, GCD-HIGHDEN-LOWC/HIGHOFF, with initial central density $\rho_c = 6 \times 10^9$ g/cm$^3$, 30/70 C/O ratio, and a high, buoyancy-driven 200 km offset radius. The abundance ratios for this model, shown as upward triangles in Figure \ref {fig:ige_ratios}, are in excellent agreement for both Ni/Fe and Mn/Fe even at $Z = 0$, while the Cr/Fe ratio is too low in comparison to SNR 3C 397. The $\chi^2$ per degree of freedom for this model, shown in Figure \ref {fig:chi2} shows poor overall agreement even at zero metallicity. Notably, unlike the lower central density models considered,  $\chi^2$ per degree of freedom increases slightly with increasing metallicity for this high-central density model, implying that the best-fit model is indeed consistent with subsolar metallicity.

\citet {yamaguchietal15} demonstrated good agreement between SNR 3C 397 Mn/Fe and Ni/Fe ratios with 1D DDT SNe models, though they required high metallicity $Z \gtrsim 5 Z_{\odot}$. Such a high metallicity is in tension with observations of galactic stellar metallicities at the galactocentric radius of 3C 397. In contrast, here we find excellent agreement for Ni/Fe and Mn/Fe for model GCD-HIGHDEN-LOWC/HIGHOFF even for $Z = 0$. We note that while the higher WD progenitor central density and lower C/O fraction may be easily incorporated into 1D models, the buoyancy-driven offset is intrinsically a multidimensional effect and requires at least a 2D simulation. The Cr/Fe ratio for GCD-HIGHDEN-LOWC/HIGHOFF  is too low in comparison to SNR 3C 397 \citep {yamaguchietal15}. Because Cr/Fe is nearly independent of $Z$, this low abundance ratio points towards a systematic effect unrelated to metallicity. Correspondingly, we next turn our attention to other explosion models.

 \subsection {Effect of a DDT and a Pure Deflagration}

We move on to address the question of the nature of the explosion mechanism in near $M_{\rm Ch}$ WDs.
While a GCD or a DDT undergoes neutronization during both deflagration and detonation phases of burning, a pure deflagration, in contrast, will of course only undergo neutronization during deflagration. However, despite the lower nuclear energy release in a pure deflagration in comparison to detonating models, the relative abundances of stable iron peak elements in the ejecta can be quite high. The reason why is simple -- although a centrally-ignited pure deflagration leaves behind a substantial portion of the WD at low densities, it burns completely through the center of the WD (Figure \ref {fig:ye}, rightmost panel), producing high central IGE yields (Figure \ref {fig:sample_lagrangians}).

In our last model variation, we consider two different explosion mechanisms. The first examines the role of a DDT, in the models DDT-HIGHDEN-LOWC/HIGHOFF and  DDT-HIGHDEN-LOWC/CENTRAL. These DDT models are identical in all respects to the GCD-HIGHDEN-LOWC/HIGHOFF model, except in its detonation mechanism. Specifically,  instead of permitting the flame bubble to break out and flow over the surface of the WD, we initiate a detonation at a transition density of $2.6\times 10^7$ g cm$^{-3}$ \citep {townsleyetal09, kruegeretal10, jacksonetal10}. We additional consider two deflagration models, DEF-STD and DEF-HIGHDEN-LOWC/CENTRAL, in which we ignite the WD centrally, and suppress the development of any detonation.

The model DEF-HIGHDEN-LOWC/CENTRAL is a pure deflagration model within the same WD progenitor structure as its GCD and DDT counterparts, but is centrally ignited, and does not undergo a detonation at any point in its evolution. The result is a failed detonation, leaving behind a significant amount of unburned carbon and oxygen fuel, and producing only a small amount of radioactive $^{56}$Ni. Its IGE abundance ratios are shown in Figure \ref {fig:ige_ratios} in the circular data points. While the Ni/Fe ratio generally falls quite close to the observed value for SNR 3C 397 across all metallicities, the Mn/Fe and Cr/Fe ratios are far in excess of the observed values. 

In contrast, the DEF-STD model, which is a centrally-ignited model within the standard progenitor WD, shows excellent overall agreement with the IGE yields in 3C 397. However, like model DEF-HIGHDEN-LOWC/CENTRAL, the DEF-STD model also produces a subenergetic, subluminous SN Ia. In contrast to previous we find good agreement with the IGE yields of SNR 3C 397 even for solar or subsolar stellar progenitor metallicity, Figure \ref {fig:chi2}. We return to this model in the discussion.

Because it detonates prior to breakout, at a higher density than model GCD-HIGHDEN-LOWC/HIGHOFF, the DDT model variants  DDT-HIGHDEN-LOWC/HIGHOFF and  DDT-HIGHDEN-LOWC/CENTRAL produce higher stable IGE abundance ratios, as shown in the square and pentagonal points in Figure \ref {fig:ige_ratios}, respectively. The higher production of stable IGE elements produces the best overall agreement with 3C 397 among all detonating models considered, as seen in Figure \ref {fig:chi2}. The best agreement is found for a progenitor with slightly super-solar metallicity, $Z = 1.5 Z_{\odot}$.

\begin{figure*}
\centering
\begin{minipage}[b]{.4\textwidth}
		\includegraphics[width=3 in]{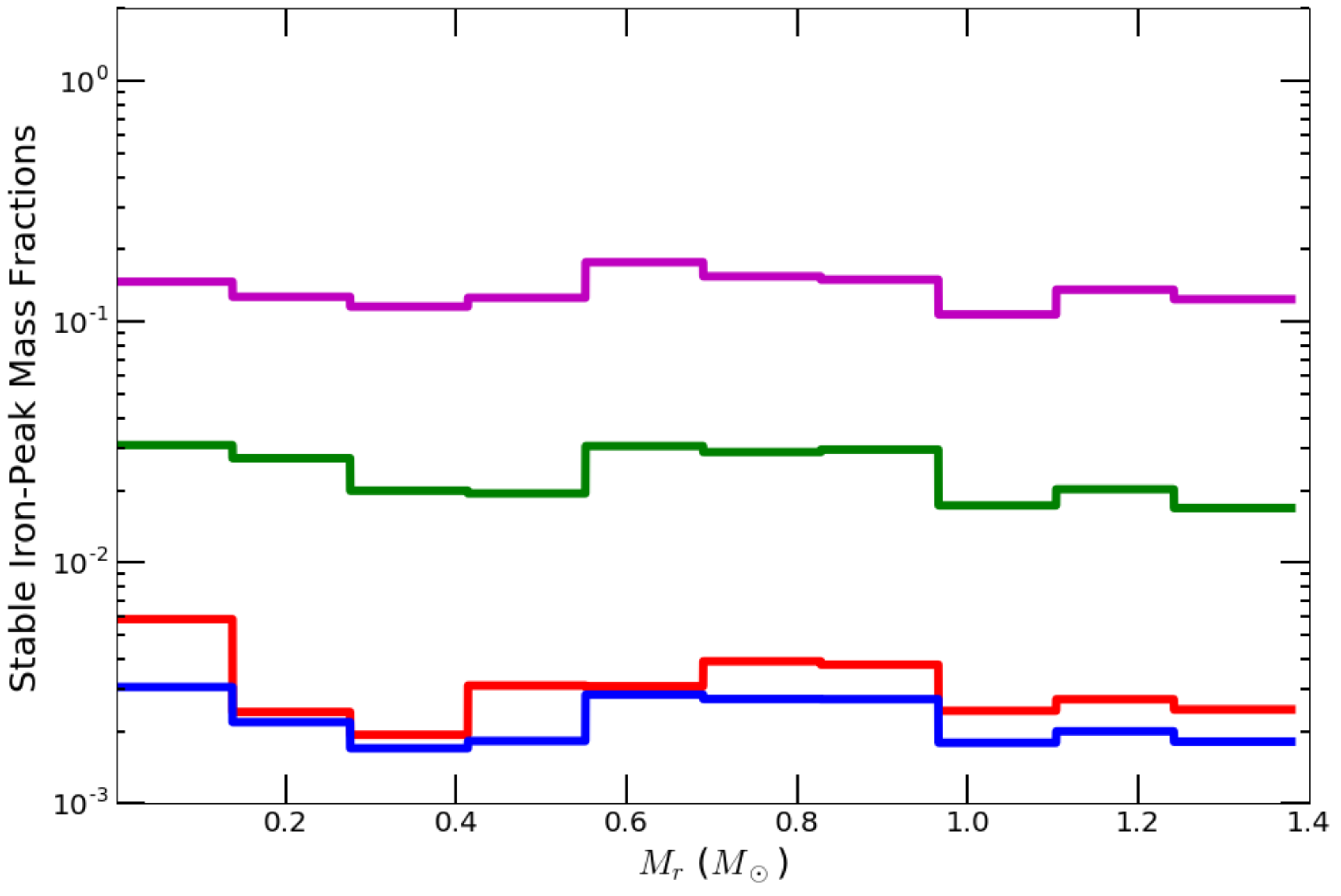}
\end{minipage}\qquad
\begin{minipage}[b]{.4\textwidth}
		\includegraphics[width=3 in]{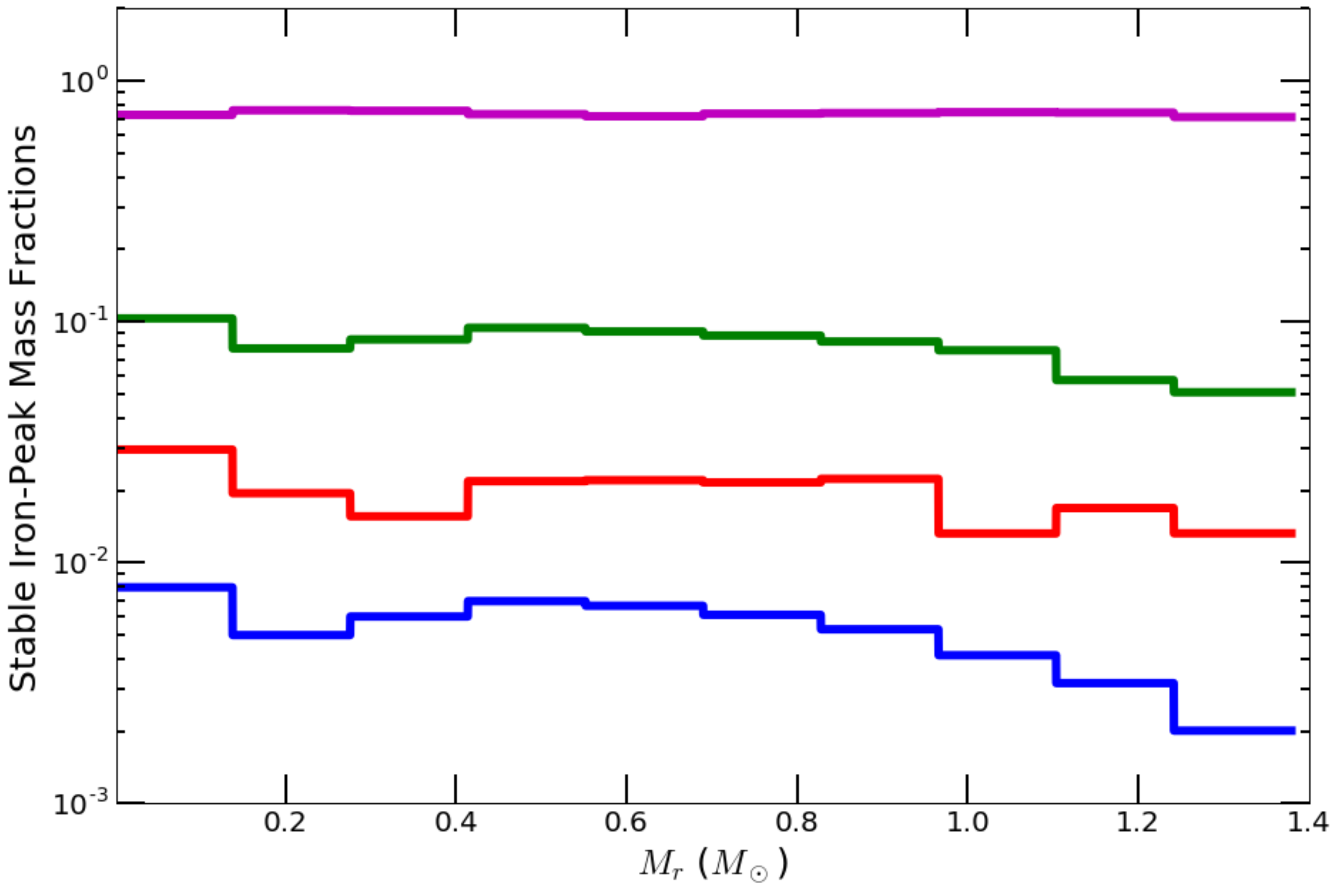}
\end{minipage}
	\caption{Stable iron peak element mass fractions plotted as a function of Lagrangian mass coordinate $M_r$, in solar masses, shown for models DEF-STD (left) and DDT-HIGHDEN-LOWC/CENTRAL (right). The purple curve represents the mass fraction of Fe in the remnant, and the red, blue, and green curves represent the mass fractions of Cr, Mn, and Ni, respectively.}\label{fig:ige_profile}
\end{figure*}

Figure \ref {fig:ige_profile} shows the key stable IGE abundance ratios for models DEF-STD and DDT-HIGHDEN-LOWC/CENTRAL, plotted as a function of Lagrangian mass coordinate $M_r$. While stable Fe is distributed nearly evenly throughout the remnant for both models,  other iron group elements are less evenly distributed. In particular, in the DDT model, Cr, Mn, and Ni are concentrated towards the center of the remnant. 


A key distinguishing characteristic between the two model variants GCD-STD and DDT-HIGHDEN-LOWC/HIGHOFF lies within the distribution of IGEs within the remnant. Because the GCD model fundamentally requires the deflagration ash to be ejected from the core of the WD and move over the surface prior to detonation, most of the burnt products of deflagration can be found far outside the core of the WD, at high velocities $\gtrsim 10^4$ km/s. In contrast, the DDT model detonates sooner, prior to the breakout of the bulk of the deflagration ashes. As a consequence, there is less trace of IGEs at high velocities in the DDT model. 


\subsection {$^{56}$Ni Nucleosynthesis}

$^{56}$Ni plays a key role in powering the optical light curve of a SN Ia, and so we focus upon its nucleosynthesis first, separately from the detailed nucleosynthetic yields of stable and decayed IGEs. Figure \ref {fig:ni56} shows the mass of $^{56}$Ni produced in each model as a function of metallicity, using the same symbols as the previous plots. Our models span a wide range of $^{56}$Ni yields, from $0.1 M_{\odot} - 1.2 M_{\odot}$. We find that the production of stable IGE with metallicity decreases the $^{56}$Ni mass by roughly 10\% in the range of $Z = 0 - 3 Z_{\odot}$, in agreement with a body of previous work -- e.g. \citet {timmesetal03} and \citet {milesetal16}. 

The $^{56}$Ni yield depends very strongly upon the WD progenitor central density as well as its C/O fraction and ignition offset. Because it leaves behind a significant amount of fuel, the DEF-HIGHDEN-LOWC/CENTRAL model has the lowest $^{56}$Ni yield of all models considered. Most of our other models are concentrated in the range of $1.0 - 1.2 M_{\odot}$ at $Z = 3 Z_{\odot}$ to $1.2 M_{\odot}$  at $Z = 0$, consistent with the expectations for single-bubble, buoyancy-driven ignitions \citep {fisherjumper15}. 

It is noteworthy that all of the detonating high-central density WD progenitor models (GCD-HIGHDEN, GCD-HIGHDEN-LOWC/HIGHOFF, DDT-HIGHDEN-LOWC/HIGHOFF, and DDT-HIGHDEN-LOWC/CENTRAL) have a range of $^{56}$Ni masses from $0.5 M_{\odot} - 0.9 M_{\odot}$, depending upon stellar progenitor metallicity, consistent with normal to bright normal SNe Ia. A large body of previous theoretical work has relied upon a vigorous deflagration phase to pre-expand the WD and produce a yield of  $^{56}$Ni typical of a normal SN Ia. Such a vigorous pre-expansion phase is challenging to achieve in  a low-central density WD progenitor with $\rho_c \sim 2 \times 10^9$ g cm$^{-3}$  with a single bubble, since the effect of buoyancy drives the WDs towards a weak deflagration and large $^{56}$Ni yield, unless the ignition radius of the bubble is within a small critical radius $\sim 20$ km \citep {fisherjumper15}. As a result, authors have typically invoked multiple ignition points to develop the requisite strong deflagration required to achieve a normal SN Ia, even though these initial conditions appear to be in tension with predictions from numerical simulations \citep {zingaleetal11, nonakaetal11,  maloneetal14}. 

Here, we find another possible way to produce $^{56}$Ni in the range $\sim 0.5 M_{\odot} - 0.9 M_{\odot}$ is through a single-bubble ignition in a high-central density WD, followed by a weak deflagration and subsequent detonation. The high-central density at detonation produces an enhancement of stable IGE, and  a $^{56}$Ni  mass consistent with a normal brightness SN Ia {\it even with a single ignition point.}  The  large nucleosynthetic yield of stable IGEs in such a low-carbon, high central density WD leads to a lower level of IMEs, implying that SNR 3C 397 may not have been a spectroscopically normal SN Ia. 

\subsection {IGE Yields in High Central Density WD Models}

We examine what IGE yields for SNR 3C 397, with a particular view towards what this may imply for the overall prevalence of SD SNe Ia, compared to the total SNe Ia rate. Pioneering work demonstrated that high central density WD progenitors yield high abundances of neutronized isotopes \citep {meyeretal96, woosley97, nomotoetal97, brachwitz00}.   We examine the isotopic abundances of all iron-peak elements in the models DEF-STD and DDT-HIGHDEN-LOWC/CENTRAL in Figure \ref {fig:isotopic_abundances}. Both models significantly overproduce a number of iron group elements, including $^{50}$Ti, $^{54}$Cr, $^{58}$Fe, and $^{62}$Ni relative to solar.  Thus, while the multi-dimensional models considered here extend the previous one-dimensional models to the current state-of-the-art,  including a more realistic multi-dimensional treatment of the flame surface, they still exhibit much the same overproduction of neutronized isotopes of the earlier models. 



These isotopic mass fractions for $^{54}$Fe, $^{52}$Cr, $^{55}$Mn, $^{58}$Ni, and  $^{60}$Ni for models DEF-STD and DDT-HIGHDEN-LOWC/CENTRAL are too large relative to solar for SNR 3C 397 to be representative of the mean yields of all SNe Ia \citep {kobayashietal06}, but cannot be excluded for a single SN Ia event. These mass fractions may in turn be suggestive of the possibility that SNR 3C 397 is representative of the class of SD SNe Ia as a whole, and that SD SNe Ia may be atypical SNe Ia events. Alternatively SNR 3C 397 may itself simply be an atypical representative of the class of SD SNe Ia. We return to this issue in the discussion.

\section {Discussion}
\label {discussion}

The nucleosynthetic yields computed from our hydrodynamic models lead to a strong preference for either the pure deflagration model DEF-STD, or the DDT model DDT-HIGHDEN-LOWC/HIGHOFF. However, previous observations of Fe K$\alpha$ emission of SNR 3C 397 indicates that the centroid energy and line luminosity from this remnant requires a bright SN Ia model, and cannot be reproduced by normal or faint SNe Ia models \citep {yamaguchietal14}. The conclusions of \citet {yamaguchietal14}  are based upon some assumptions, including a uniform ambient medium density. However,  for a range of reasonable ambient medium densities, subluminous SNe Ia models of \citet {yamaguchietal14} are inconsistent with the centroid and line luminosity data, and therefore make a subluminous SNe such as our DEF-STD model an unlikely possibility for SNR 3C 397. Consequently, the DDT model DDT-HIGHDEN-LOWC/HIGHOFF is most consistent with both the IGE yields and Fe K$\alpha$ emission data.

Assuming the SD channel to be the dominant contributor to the SNe Ia population, \citet {woosley97} and \citet {nomotoetal97}  suggested the central WD density at ignition must be $\lesssim$ 2$\times$10$^9$ g cm$^{-3}$ in order to avoid overproducing key neutron-rich isotopes such as $^{54}$Cr and $^{50}$Ti. Additional models including improved electron capture rates  \citep{brachwitzetal00} arrived at a similar conclusion. In turn, such a  central density at ignition implies relatively rapid accretion rates of $\dot {M} \gtrsim 10^{-7} M_{\odot}$ yr$^{-1}$.
Such high accretion rates would be consistent with supersoft X-ray sources. Consequently, requiring SDs to be the dominant SNe Ia channel logically demands sufficiently numerous supersoft X-ray sources in order to explain the SNe Ia rate, in well-known conflict with observation -- see e.g. \citet {distefano10}. 

However, recent advances favor both double-degenerate and SD SNe Ia occurring in nature, with the SD channel being subdominant. This fundamental reshaping of the basic picture for SNe Ia has important ramifications for SD progenitors. Significantly,  the SD population is likely to be a small fraction of the total SNe Ia rate $\lesssim 10 - 20\%$ \citep {haydenetal10, biancoetal11, chomiuketal16}, which reduces the overproduction problem of neutron-rich stable IGEs by a simple reduction of the overall rate of the SD channel. \added {SNR 3C 397 has the highest stable IGE yields of any SNR, though this may partially reflect the uniquely deep observations of Suzaku in this older remnant \citep {yamaguchietal15}. It is possible, however, that it is representative of the broader class of SD SNe Ia.} Thus the problems associated with overproduction of isotopes such as $^{54}$Cr, $^{50}$Ti, $^{58}$Fe, and $^{62}$Ni  may not be as significant as previously thought, which in turn implies that high-central density progenitor WDs may constitute a significant fraction of all SD SNe Ia.


Furthermore, another implication of high central density WDs, and the  necessarily lower accretion rates involved, is associated with the problem of supersoft X-ray sources \citep {distefano10}. If the population of high central density WD progenitors of SD SNe Ia is indeed significant, this may help account for the observed deficiency of super-soft X-ray sources as SNe Ia progenitors. Lower accretion rates have commonly been associated with the ejection of material based on the abundances observed in ejecta from hydrogen shell flashes, i.e. novae \citep {gehrzetal98}, and the similarity of ejected and accreted masses \citep {townsleybildsten04}.  The issue of the accreted mass retention fraction is, however, by no means a  settled question -- see e.g. \citet {hillmanetal16}. Mass gain may be possible at lower accretion rates than stable burning, especially in wider systems with a larger Roche lobe.




Several authors have explored the possibility of using the equivalent width ratio of Cr to Fe K$\alpha$ lines in X-ray observations of young remnants as probes of the SNR type and of the explosion physics \citep {yangetal09, yangetal13}. In these young remnants, however, the reverse shock has not yet propagated to the center of the remnant as in SNR 3C 397. These authors find that Cr and Fe are well-mixed throughout the outer layers of young SNRs, and argue that the measured flux ratios of the Cr and Fe K$\alpha$ lines accurately capture the global mass ratios. Figure \ref {fig:ige_profile} shows the IGE mass ratios of the outer portions of the SNR are indeed more uniform than the deep interior. However, as we also show in  Figure \ref {fig:ige_profile}, the IGE yields are far from well-mixed throughout the SNR. In particular, the IGEs are concentrated towards the center of the SNR in a DDT, and caution must be applied when directly comparing observed equivalent widths in young SNRs against spatially-integrated model abundances. Detailed future multidimensional models of realistic young SNRs may provide predictions for X-ray observations, which may help resolve the question of their stellar progenitors as well.
      
The highest central density $\rho_c \sim 5 \times 10^9$ g cm$^{-3}$ models considered by \citet {lesaffreetal06}  result from larger initial WD mass near the maximum C/O WD mass of 1.2 $M_{\odot}$. Because of the enhancement of C burning due to electron screening at higher densities, the ignition curve is nearly vertical in the $\rho-T$ plane, leading to a convergence in the central density for a range of initial WD masses (their Figure 4). Their Figure 11 shows that the distribution of central densities becomes increasingly peaked around the highest values at later delay times ($> 0.8$ Gyr). The size of the convective core depends on the details of the ignition, and is $\sim 0.85 - 0.96\ M_{\odot}$ for the highest central density cases considered. In the convective core at ignition, their carbon mass fraction is $\sim 0.27 - 0.28$ for their highest central density progenitors, similar to the values which we find best match SNR 3C 397.

Previous authors \citep {kruegeretal10, kruegeretal12, seitenzahletal11, seitenzahletal13b} have explored the effect of varying the central density of the progenitor WD, though while adopting differing ignitions and differing flame models, and differing DDT transition conditions. While these authors agreed that the total amount of IGEs (both stable and unstable) generally increase with increasing WD central density, they reached somewhat distinct conclusions regarding the $^{56}$Ni nucleosynthetic yield and the production of stable IGEs. In particular, while \citet {kruegeretal10, kruegeretal12} found that the yields of stable IGEs increased with increasing WD central density, while $^{56}$Ni decreased.  The total production of all IGEs ($^{56}$Ni plus stable IGEs) was roughly constant with respect to variation of the central density in these models.  In contrast,  \citet {seitenzahletal11, seitenzahletal13b} concluded that the central density played the role of a secondary parameter, with their multipoint ignition having a more significant impact upon the outcome of their models.  In the work of \citet {seitenzahletal11, seitenzahletal13b}, the overall IGE yield also varied with central density, leading to a more complex behavior such that the $^{56}$Ni yield increased with higher central density. The reasons for these differences are summarized in both \citet {seitenzahletal11} and \citet {kruegeretal12}; here we note that one important distinction between these two sets of models was the choice of a perturbed central ignition in the case of \citet {kruegeretal10, kruegeretal12} and off-centered multipoint ignitions in \citet {seitenzahletal11, seitenzahletal13b}. Because the deflagration phase is sensitive to the ignition, in a set of multi-point ignition models in which the ignition parameters are highly varied, the choice of ignition will itself tend to dominate the effects of WD progenitor metallicity and chemical composition. Furthermore, while \citet {kruegeretal10, kruegeretal12} studied multiple realizations of the ignition condition at each central density and took the average of these, \citet {seitenzahletal11, seitenzahletal13b} only studied one realization at each central density. In this sense \citet {kruegeretal10, kruegeretal12} also find that the ignition distribution is the primary parameter, but they average over different ignition distributions in order to measure the systematic effect of the central density.

Extensive surveys for radio emission associated with SNe Ia have so far found non-detections in all sources considered. 
\citet {chomiuketal16} place limits on the mass accretion rates in the range $\dot{M} <  10^{-9} - 10^{-4} M_{\odot}$ yr$^{-1}$. We note that for standard accretion models for high central density WDs $\rho_c > 5 \times 10^9$ g cm$^{-3}$, the implied mass accretion rates $\dot{M} \simeq 10^{-9} M_{\odot}$ yr$^{-1}$ \citep {nomotoetal84} are beneath the derived radio lower limits. Thus the current radio data does not exclude a potential population of slowly-accreting WDs which ignite as SD SNe Ia at high central densities comparable to that we infer for 3C 397. However, the radio lower bounds on the accretion rates are nearing even those values expected for high-central density SD SNe Ia, which suggests that future radio observations should be able to either detect the circumstellar material in these systems, or strongly exclude these as progenitors of SNe Ia \citep {chakrabortietal16}.

\section {Conclusions}
\label {conclusions}

To summarize our key conclusions :

\begin {enumerate}

\item Stable iron peak element abundances of 3C 397 are consistent with either a centrally-ignited standard white dwarf progenitor undergoing a pure deflagration (model DEF-STD), or a  high-central density, low-carbon C/O progenitor undergoing a weak deflagration energy release followed by a DDT (model DDT-HIGHDEN-LOWC/HIGHOFF). Because the X-ray observations of the centroid energies and luminosities of 3C 397 make a subluminous event unlikely \citep {yamaguchietal14}, our  high-central density detonating model DDT-HIGHDEN-LOWC/HIGHOFF is most consistent with the energetic and nucleosynthetic constraints provided by the observations. 

\item High-central density WDs imply a very low $\dot {M} \simeq 10^{-9} M_{\odot}$ yr$^{-1}$. Such low rates are in the nova regime, and are supportive of some recent findings that symbiotic novae may continue to grow in mass -- see e.g.  \citet {starrfieldetal12}. Alternatively, such a low effective rate  may be indicative of a low, but non-negligible retention fraction of C/O from accreted material. 


\item With their lower accretion rates, high central density WD progenitors naturally predict very low winds, consistent with derived bounds on radio emission in SNe Ia.

\end {enumerate}

\added {Our conclusions rest upon a number of assumptions. In particular, all simulations presented here have assumed  near-$M_{\rm Ch}$ WD progenitors in 2D axisymmetry, while a full 3D geometry permits enhanced burning \citep {jordanetal08}. Further, while {\it ab-initio} numerical simulations performed to date demonstrate single-point ignitions  \citep {zingaleetal11, nonakaetal11,  maloneetal14}, these simulations have relied upon a single turbulent realization within a single WD progenitor, so the issue has not been laid fully to rest. There are additional uncertainties in the subgrid burning model; our adopted model typically burns less in the deflagration phase than other models \citep {jordanetal08}.  However, all of these assumptions lead to higher deflagration energy release than the models considered here. As a consequence, the requirement to achieve significant burning at high densities in order to match SNR 3C 397 would generally push progenitor WD models  including either 3D geometry, multiple point ignitions, or alternative subgrid burning models to even higher central densities than those found here. Lastly, strong compressions generated by the inwardly-moving shock in sub-$M_{\rm Ch}$ WD  double-detonation models may also produce significant neutronization, although previous work \citep {yamaguchietal15} suggests their stable IGE yields are inconsistent with 3C 397 for near-solar metallicity.}

\added {However, our conclusions could be significantly impacted by significant neutronization during the simmering phase, if they are larger than predicted by current models  \citep {martinezrodriguezetal16}. New data on the 3C 397 Ca/S ratio \citep {martinezrodriguezetal17}, compared against a new suite of 3D simulations building and extending upon this work may help to either rule out or support the models presented here.}






{\bf Acknowledgements}  The authors thank Stan Woosley, Hiroya Yamaguchi, Eduardo Bravo, Carles Badenes, Or Graur,  Brad Meyer, and Chris Byrohl for informative discussions. FXT was supported by NASA under the Theoretical and Computational Astrophysics Networks (TCAN) grant NNX14AB53G, by NSF under the Software Infrastructure for Sustained Innovation (SI$^2$) grant 1339600 and grant PHY 08-022648 for the Physics Frontier Center ``Joint Institute for Nuclear Astrophysics - Center for the Evolution of the Elements (JINA-CEE),'' and the Simons Foundation. RTF thanks the Institute for Theory and Computation at the Harvard-Smithsonian Center for Astrophysics, and the Kavli Institute for Theoretical Physics, supported in part by the national Science Foundation under grant NSF PHY11-25915, for visiting support during which this work was completed.

\software {We utilize the adaptive mesh refinement code FLASH 4.0.1 with additional units for SNe Ia described most recently by \citet {townsleyetal16}, and modules available from \url {http://astronomy.ua.edu/townsley/code}. For nucleosynthetic post-processing we use TORCH, as described by \citet{timmes99}, and available from \url {http://cococubed.asu.edu/code_pages/net_torch.shtml}. For plotting and analysis, we have made use of yt \citep {Turk_2011}, \url {http://yt-project.org/}.}

\bibliography{converted_to_latex}


\listofchanges

\end{document}